\documentclass[12pt]{article}
\usepackage[usenames,dvipsnames,svgnames,table]{xcolor} 
\usepackage{kotex}
\usepackage{graphicx}
\usepackage{epsfig}
\usepackage{amssymb, amsmath, mathtools}
\usepackage{verbatim}
\usepackage{natbib}
\usepackage[affil-it]{authblk}
\usepackage{mathrsfs}
\usepackage{here}
\usepackage{makecell}
\usepackage{tikz}
\usepackage{enumerate}
\usepackage[shortlabels]{enumitem}
\usepackage{yfonts}
\usepackage{comment}
\usepackage{longtable}
\usepackage{tikz,amsmath, amssymb,bm,color}
\usetikzlibrary{circuits.logic.US,circuits.logic.IEC,fit}
\usepackage{algorithm2e}
\usepackage{multirow}
\usepackage{physics}
\usepackage{amsthm}
\usepackage{makecell}
\usepackage{ulem}
\usepackage{proofread}
\usepackage{booktabs}
\usepackage{array}
\usepackage{xr}
\usepackage{adjustbox}
\usepackage{lineno}
\usepackage{import}
\usepackage{subcaption}

\usepackage[title,titletoc]{appendix}
\usepackage[colorlinks]{hyperref} 
\usepackage[colorinlistoftodos, textsize=scriptsize]{todonotes} 

\usepackage[framemethod=TikZ]{mdframed}
\mdfdefinestyle{JL}{%
	linecolor=blue,
	outerlinewidth=1pt,
	roundcorner=2pt,
	innertopmargin=0.1\baselineskip,
	innerbottommargin=0.1\baselineskip,
	innerrightmargin=2pt,
	innerleftmargin=2pt,
	backgroundcolor=orange!100!white}

\hypersetup{
	colorlinks,
	citecolor=Black,
	linkcolor=Red,
	urlcolor=Blue}

\includecomment{note} 

\newtheorem{theorem}{Theorem}[section]
\newtheorem{lemma}[theorem]{Lemma}
\newtheorem{definition}[theorem]{Definition}
\newtheorem{proposition}[theorem]{Proposition}
\newtheorem{corollary}[theorem]{Corollary}

\newtheorem{remark}[theorem]{Remark}

\everymath{\displaystyle}

\newcommand{\mc}{\mathcal}

\newcommand{\be}{\begin{equation}}
	\newcommand{\ee}{\end{equation}}
\newcommand{\bea}{\begin{eqnarray*}}
	\newcommand{\eea}{\end{eqnarray*}}
\newcommand{\bean}{\begin{eqnarray}}
	\newcommand{\eean}{\end{eqnarray}}
\newcommand{\ben}{\begin{enumerate}}
	\newcommand{\een}{\end{enumerate}}
\newcommand{\bi}{\begin{itemize}}
	\newcommand{\ei}{\end{itemize}}
\newcommand{\brem}{\begin{remark}}
	\newcommand{\erem}{\end{remark}}
\newcommand{\bcen}{\begin{center}}
	\newcommand{\ecen}{\end{center}}
\newcommand{\bsv}{\begin{semiverbatim}}
	\newcommand{\esv}{\end{semiverbatim}}

\newcommand{\bt}{\begin{theorem}}
	\newcommand{\et}{\end{theorem}}
\newcommand{\bl}{\begin{lemma}}
	\newcommand{\el}{\end{lemma}}
\newcommand{\bd}{\begin{definition}}
	\newcommand{\ed}{\end{definition}}
\newcommand{\bc}{\begin{corollary}}
	\newcommand{\ec}{\end{corollary}}
\newcommand{\bp}{\begin{proposition}}
	\newcommand{\ep}{\end{proposition}}

\newcommand{\bff}{ \mathbf{f}}

\newcommand{\bfy}{ \mathbf{y}}

\newcommand{\bbE}{{ \mathbb{E}}}

\usepackage{xr}
\makeatletter
\newcommand*{\addFileDependency}[1]{
	\typeout{(#1)}
	\@addtofilelist{#1}
	\IfFileExists{#1}{}{\typeout{No file #1.}}
}
\makeatother

\modulolinenumbers[5]

\usepackage[textwidth=16cm, textheight=22cm, centering]{geometry}

\title{Bayesian Node-Level Outlier Detection for Graph Signals}
\author[1]{Seongmin Kim$^{\dagger}$}
\author[2]{Kyusoon Kim$^{\dagger,*}$}

\affil[1]{Human-Centered Artificial Intelligence Research Institute, Ewha Womans University, Seoul 03760, Korea}
\affil[2]{Department of Statistics and Actuarial Science, Soongsil University, Seoul 06978, Korea}

\allowdisplaybreaks
\begin{document}
	
\maketitle

\footnotetext[2]{These authors contributed equally to this work.}
\footnotetext[1]{Corresponding author: kyusoon.kim@ssu.ac.kr}

\begin{abstract}
This paper proposes a fully Bayesian framework for node-level outlier detection in graph signals, where measurements are observed on the nodes of an underlying graph. Unlike traditional outlier detection methods, our approach accounts for the relational dependencies induced by the graph, identifying outliers that disrupt the underlying smoothness. We model the observed signal as a combination of a graph-smooth component, captured via an intrinsic Gaussian Markov random field (IGMRF) prior, and a sparse outlier component modeled by a spike-and-slab prior. A key advantage of the proposed method is its ability to provide principled uncertainty quantification by estimating the posterior probability that each node is an outlier, rather than enforcing a deterministic binary decision. To facilitate posterior inference, we develop an efficient Gibbs sampling algorithm. We demonstrate the effectiveness of the proposed method through simulation studies on various graph structures, as well as a real data analysis of PM2.5 levels in California, exploring their relationship with wildfire occurrences.
\end{abstract}

\noindent {\it \bf Keywords}: Bayesian inference; Graph signal; Outlier detection; Uncertainty quantification.   

\section{Introduction} 
In recent years, the rapid development of information technology has led to a substantial increase in network-structured data. Many modern datasets are naturally represented as graphs (or networks), where complex relational structures arise in social networks, sensor networks, transportation systems, biological networks, and economic systems. In such settings, measurements are often collected at the nodes of a graph. A graph signal refers to the data observed on the nodes of a graph, that is, a collection of values indexed by the node set \citep{shuman2013emerging}. Throughout the paper, we use the terms graph signal and graph-indexed data interchangeably. Examples include traffic flow measured at intersections, temperatures recorded at spatially distributed sensors, and activity levels observed on social networks. As graph-indexed data become increasingly common, statistical modeling and inference for such data have attracted growing attention \citep{leus2023graph}.

Outlier detection plays a fundamental role in the analysis of graph signals. In many applications, abnormal node-level observations may indicate sensor malfunction, cyber-attacks, fraudulent activity, structural failures, or rare but critical events. Failure to properly identify such anomalous signals can lead to biased estimation, distorted inference, and unreliable decision-making. Unlike classical outlier detection problems in independent data settings, graph signals exhibit dependency structures induced by the underlying graph topology. This relational structure implies that abnormal behavior is often characterized not only by extreme magnitude but also by inconsistency with neighboring nodes. In this context, normal behavior of signal values on a graph is typically defined through a smoothness assumption, where adjacent nodes are expected to exhibit similar values. Therefore, developing principled methods for outlier detection that account for graph dependence is both practically important and methodologically challenging.

A Bayesian framework provides a natural approach for addressing these challenges. 
It enables the incorporation of structural assumptions through prior distributions while simultaneously estimating latent indicators that represent abnormal nodes. 
Moreover, posterior distributions offer a principled way to quantify uncertainty in both model parameters and outlier probabilities. 
In graph-indexed data, outlier identification is inherently challenging, as a node may not be globally extreme but can still be considered an outlier if it exhibits behavior that is inconsistent with its neighboring nodes. 
This challenge can be partially alleviated by representing uncertainty through posterior probabilities of being outliers, which provide richer information than a hard binary classification.

The primary contribution of this paper is the development of a fully Bayesian model for detecting outlying nodes in graph signals. The proposed framework enables principled uncertainty quantification and, importantly, provides the posterior probability that each node is an outlier. Furthermore, we develop an efficient Gibbs sampling algorithm to facilitate posterior inference. Through numerical experiments on both simulated and real-world datasets, we demonstrate that the proposed approach achieves strong performance compared to existing methods.

The remainder of this paper is organized as follows. Section~\ref{sec:preliminary} introduces the fundamental notation and reviews key concepts in graph signal processing. Section~\ref{sec:main_results} presents the proposed Bayesian model along with the corresponding Gibbs sampling procedure. Sections~\ref{sec:simul_studies} and \ref{sec:real_data} provide experimental results based on simulation studies and real-world data, respectively. Section~\ref{sec:conclusion} concludes the paper with a discussion of findings and directions for future research. The \textsc{R} codes are available at \url{https://github.com/qsoon/graph-bayes-outlier}.


\section{Preliminaries}\label{sec:preliminary}
\subsection{Basics of Graph Signal Processing} \label{sec:notation}
Denote by $\mc{G} = (\mc{V}, \mc{E}, \mathbf{W})$ an undirected weighted graph, where $\mc{V}$ is the set of $N$ nodes (i.e., $|\mc{V}| = N$), $\mc{E}$ is the set of edges with nonnegative weights, and $\mathbf{W} = (w_{ij})_{N \times N}$ is the corresponding weighted adjacency matrix. The entry $w_{ij}$ represents the strength of the connection between nodes $i$ and $j$, with larger values indicating stronger associations and $w_{ij} = 0$ indicating no edge between the two nodes.

The (unnormalized) graph Laplacian is defined as $\mathbf{L} = \mathbf{D} - \mathbf{W}$, where $\mathbf{D} = \operatorname{diag}(d_1, \dots, d_N)$ is the degree matrix with $d_i = \sum_{j=1}^N w_{ij}$. Since $\mathbf{L}$ is symmetric and positive semi-definite, it admits the eigendecomposition $\mathbf{L} = \mathbf{U} \mathbf{\Lambda} \mathbf{U}^\top$, where the columns of $\mathbf{U}$ are orthonormal eigenvectors of $\mathbf{L}$ and $\mathbf{\Lambda}$ is a diagonal matrix of nonnegative eigenvalues. For a connected graph, the graph Laplacian $\mathbf{L}$ has $\mathbf{1}$ as an eigenvector associated with eigenvalue 0, implying that $\mathbf{L}$ is singular.

\begin{figure}[!t]
    \centering
    \includegraphics[width=0.5\linewidth]{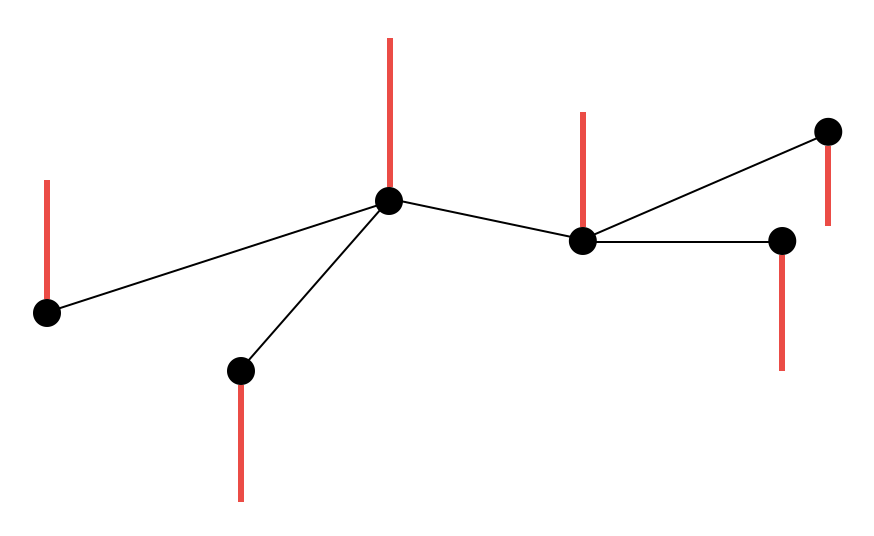}
    \caption{An example of a graph signal.}
    \label{fig:graphsignal_ex}
\end{figure}

A graph signal refers to data observed on the nodes of a graph. 
Restricting attention to real-valued data, 
a graph signal $\mathbf{y}$ on $\mc{G}$ can be represented 
as a vector $(y_1, \ldots, y_N)^\top \in \mathbb{R}^N$, 
which may be viewed as a function from $\mc{V}$ to $\mathbb{R}$. An example of a graph signal is illustrated in Figure~\ref{fig:graphsignal_ex}. In graph signal processing, the smoothness of $\mathbf{y}$ is commonly quantified by the graph Laplacian quadratic form:
\[
\mathbf{y}^\top \mathbf{L} \mathbf{y} = \frac{1}{2}\sum_{i,j=1}^N w_{ij}(y_i - y_j)^2.
\]
This value is small when the signal values at adjacent nodes with large edge weights are similar, indicating global smoothness over the graph.

Given the eigendecomposition $\mathbf{L} = \mathbf{U} \mathbf{\Lambda} \mathbf{U}^\top$, 
the graph Fourier transform (GFT) of a signal $\mathbf{y}$ is defined as
\[
\tilde{\mathbf{y}} = \mathbf{U}^\top \mathbf{y},
\]
which maps the signal from the node domain to the graph spectral domain, yielding the graph Fourier coefficients.
The inverse transform is given by $\mathbf{y} = \mathbf{U}\tilde{\mathbf{y}}$, which reconstructs the original signal from its graph Fourier coefficients \citep{sandryhaila2014discrete}.

\subsection{Related Work} \label{sec:relatedwork}
Several approaches have been proposed for outlier detection in graph signals. \cite{sandryhaila2014discrete} proposed a framework based on the graph Fourier transform \citep{sandryhaila2013discrete} and high-pass graph filtering to detect graph signals in which measurements at certain nodes are corrupted or exhibit abnormal behavior. 
\cite{egilmez2014spectral} introduced a spectral anomaly detection framework that leverages spectral decomposition using graph-based filtering or principal component analysis.  \cite{gopalakrishnan2019identification} proposed an outlier detection method based on the total variation metric defined on graphs.
However, these approaches primarily aim to identify anomalous graph signals at the level of the entire network rather than pinpointing which specific nodes contain corrupted signal values. That is, they aim to detect globally anomalous signals, rather than performing node-level outlier detection. Consequently, these methods are not fully aligned with our objective of identifying corrupted nodes within a single observed graph signal.

More closely related to our goal of node-level outlier detection, \cite{lewenfus2019use} proposed an outlier detection approach that extracts features based on vertex-frequency analysis (VFA) methods, such as the windowed graph Fourier transform \citep{shuman2012windowed} and the spectral graph wavelet transform \citep{hammond2011wavelets}, and feeds them into the local outlier factor algorithm \citep{breunig2000lof}. 
\cite{francisquini2022community} proposed a community-based outlier detection algorithm using spectral graph filtering.
\cite{qin2026adaptive} modeled the observed graph signal as the sum of a graph-smooth normal component and a node-sparse anomaly component, and simultaneously learned the graph structure and identified outlying nodes via an augmented Lagrangian optimization framework.

Although these methods operate at the node level, they are primarily developed within deterministic or optimization-based frameworks. As a result, they typically rely on tuning parameters whose selection may substantially affect performance, and they do not explicitly quantify uncertainty in the identification of corrupted nodes.


In contrast, the Bayesian literature provides a principled framework for outlier detection with uncertainty quantification. 
\cite{box1968bayesian} proposed a Bayesian linear model for identifying outliers in regression settings, while \cite{george1993variable} introduced a model using a Gaussian mixture prior along with a Gibbs sampling procedure. 
\cite{justel2001bayesian} addressed the masking effect that arises when multiple outliers are present and proposed a method based on classification variables. 
More recently, \cite{silva2019bayesian} applied Bayesian models to outlier detection in time series data.

Bayesian approaches are particularly appealing in our setting, as they enable posterior inference and provide probabilistic measures of uncertainty. In particular, they allow one to estimate the probability that each node is corrupted, while also quantifying uncertainty in model parameters.

Despite these advantages, existing Bayesian methods have not been specifically developed for node-level outlier detection in graph signals. This motivates the Bayesian framework proposed in this paper, which aims to identify corrupted nodes while simultaneously quantifying posterior uncertainty.

\section{Bayesian Outlier Detection}\label{sec:main_results}
\subsection{Proposed Model}\label{subsec:model}
We consider a graph signal $\mathbf{y}$ observed on a connected graph $\mathcal{G} = (\mathcal{V}, \mathcal{E}, \mathbf{W})$ with $N$ nodes. We assume that the underlying signal varies smoothly over the graph, while the outlier component is sparse. Under this assumption, we model the observed signal as
\begin{equation*} \label{eq:proposed_model}
\mathbf{y} = \mathbf{f} + \mathbf{s} \odot \boldsymbol{\delta} + \boldsymbol{\epsilon},
\qquad
\boldsymbol{\epsilon} \sim \mathcal{N}(\mathbf{0}, \tau^{-1}\mathbf{I}_N),
\end{equation*}
where $\mathbf{f}=(f_1,\ldots,f_N)^\top \in \mathbb{R}^N$ denotes the underlying signal, $\mathbf{s}=(s_1,\ldots,s_N)^\top\in\{0,1\}^N$ is a binary indicator vector for outliers, $\boldsymbol{\delta}=(\delta_1,\ldots,\delta_N)^\top\in\mathbb{R}^N$ represents the magnitudes of the corresponding outlier effects, and $\tau>0$ is the noise precision parameter. Here, $\odot$ denotes the element-wise (Hadamard) product.

To encode smoothness of $\mathbf{f}$, we assign an intrinsic Gaussian Markov random field (IGMRF) prior \citep{rue2005gaussian} defined through the graph Laplacian:
\begin{equation*} \label{eq:igmrf_prior}
\pi(\mathbf{f} \mid \gamma)
\propto
\gamma^{r/2}
\exp\left(
-\frac{\gamma}{2}\mathbf{f}^\top \mathbf{L}\mathbf{f}
\right),
\end{equation*}
where $r=\mathrm{rank}(\mathbf{L})$ and $\gamma > 0$ is a precision parameter controlling the degree of smoothness. Since $\mathbf{f}^\top \mathbf{L}\mathbf{f}$ quantifies variation across neighboring nodes, larger values of $\gamma$ place more prior mass on smoother graph signals. For a connected graph, the graph Laplacian $\mathbf{L}$ is singular, and its null space is spanned by the constant vector $\mathbf{1}$. Therefore, this IGMRF prior is intrinsic and is defined on the subspace orthogonal to $\mathbf{1}$.

For the sparse outlier component, we employ a spike-and-slab prior by introducing a binary indicator $s_i$ and a latent Gaussian slab variable $\delta_i$ for each node $i = 1, \dots, N$:
\begin{equation*} \label{eq:spike_and_slab}
s_i \mid \pi_i \sim \operatorname{Bernoulli}(\pi_i), \quad
\pi_i \sim \operatorname{Beta}(\alpha, \beta), \quad
\delta_i \sim \mathcal{N}(0, \tau_\delta^{-1}).
\end{equation*}
Here, $\pi_i \in [0,1]$ represents the node-specific outlier probability, which determines the prior probability of node $i$ being an outlier. The outlier contribution at node $i$ is modeled as the product $s_i \delta_i$. Equivalently, the induced prior on the node-specific outlier effect $\beta_i = s_i \delta_i$ has a spike-and-slab form:
\begin{equation*} \label{eq:induced_prior}
\beta_i\mid \pi_i \sim (1 - \pi_i) \delta_0 + \pi_i \mathcal{N}(0, \tau_\delta^{-1}),
\end{equation*}
where $\delta_0$ denotes the Dirac point mass at zero. This prior encourages sparsity by allowing $\beta_i$ to be exactly zero when $s_i = 0$, while providing a flexible Gaussian slab for modeling the magnitude of nonzero outliers when $s_i = 1$.

Finally, we complete the hierarchical model by assigning
\begin{equation*}
\pi(\tau) \propto \tau^{-3/2},
\qquad
\pi(\gamma) \propto \gamma^{-3/2}.
\end{equation*}
The motivation for these priors and the corresponding posterior propriety are discussed in Section~\ref{subsec:prior}.

The resulting joint posterior distribution is given by
\begin{align*}
\pi\!\left(
\mathbf{f},\mathbf{s},\boldsymbol{\delta},\boldsymbol{\pi},\tau,\gamma
\mid
\mathbf{y}
\right)
\propto\;
&
\left[
\mathcal{N}\!\left(
\mathbf{y} \mid \mathbf{f} + \mathbf{s}\odot\boldsymbol{\delta},\tau^{-1}\mathbf{I}_N
\right)
\pi(\mathbf{f} \mid \gamma)
\right]
\nonumber\\
&\times
\prod_{i=1}^{N}
\left[
\pi(s_i \mid \pi_i)\,\pi(\pi_i)\,\pi(\delta_i )
\right]
\pi(\tau)\pi(\gamma).
\end{align*}

\subsection{Prior Specification and Hyperparameter Choice}\label{subsec:prior}
For each node-specific outlier probability $\pi_i$, we assign a Beta prior: 
$$\pi_i\sim Beta(\alpha,\beta).$$
To reflect the prior knowledge that outliers are rare occurrences in the graph, we set the hyperparameters to $\alpha=1$ and $\beta=9$. This configuration results in a prior mean of
$$\bbE[\pi_i] = \dfrac{\alpha}{\alpha+\beta}=0.1.$$

For the precision parameters $\tau$ and $\gamma$, we assign the priors:
\begin{equation*}
\pi(\tau) \propto \tau^{-3/2},
\qquad
\pi(\gamma) \propto \gamma^{-3/2}.
\end{equation*}
These priors are equivalent to uniform priors on the corresponding standard deviations. They belong to the class of weakly informative priors discussed by \cite{gelman2006prior} for the variance parameters in hierarchical models. Since these priors are improper, it is necessary to verify that the posterior distribution is proper.

\begin{theorem}[Posterior Propriety]\label{thm:post_proper}
Assume the graph $\mathcal{G}$ is connected, so that 
$$\mathrm{rank}(\mathbf L)=r=N-1.$$
Suppose that $N\ge 3$,  $\alpha>0$, $\beta>0$, and $\tau_\delta>0$. Then, for Lebesgue-almost every observed signal $\mathbf{y} \in \mathbb{R}^N$, the posterior distribution $\pi(\mathbf{f},\mathbf{s},\boldsymbol{\delta},\boldsymbol{\pi},\tau,\gamma \mid \mathbf{y})$ associated with the hierarchical model in Section~\ref{subsec:model} is proper.
\end{theorem}

\begin{proof}
    The proof is provided in Appendix~\ref{sec:proofs}.
\end{proof}

The parameter $\tau_\delta$ controls the prior scale of the outlier effects. Since $\delta_i \sim \mathcal{N}(0,\tau_\delta^{-1})$, larger values of $\tau_\delta$ induce stronger shrinkage toward zero, thereby favoring smaller nonzero outlier effects. To calibrate this scale in a data-dependent manner, one natural choice is
$$\tau_\delta = \frac{1}{2\,{\operatorname{Var}}(\mathbf{y})}.$$
This sets the slab variance according to the overall variability of the observed graph signal. However, because $\operatorname{Var}(\mathbf{y})$ is sensitive to outliers, severe outliers can artificially inflate this quantity and lead to an unstable choice of $\tau_\delta$. To obtain a more robust scale estimate, we instead use the median absolute deviation (MAD), defined by
$$\operatorname{MAD}(\mathbf{y}) = \operatorname{med}\bigl(|\mathbf{y}-\operatorname{med}(\mathbf{y})|\bigr),$$
where $\operatorname{med}(\mathbf{y})$ denotes the median of $\mathbf{y}$. Under normality, $k\times\operatorname{MAD}(\mathbf{y})$ is a consistent estimator of the standard deviation, where
$$k=\frac{1}{\Phi^{-1}(3/4)},$$
and $\Phi$ denotes the cumulative distribution function of the standard normal distribution. Therefore, we set
$$\tau_\delta
=\frac{1}{2k^2\,{\operatorname{MAD}}(\mathbf{y})^2}.
$$

\subsection{Posterior Inference via Gibbs Sampling}

We use a Gibbs sampler to draw samples from the posterior distribution. Let $\mathbf{L}=\mathbf{U}\boldsymbol{\Lambda}\mathbf{U}^\top$ be the eigendecomposition of the graph Laplacian and define the effective observation $\mathbf{y}^{\mathrm{eff}}=\mathbf{y}-\mathbf{s}\odot\boldsymbol{\delta}$. Throughout the paper, $\mathrm{Gamma}(a,b)$ denotes the Gamma distribution with shape $a$ and rate $b$.

The Gibbs sampling algorithm iteratively updates each parameter by drawing from its full conditional distribution. Let $\boldsymbol{\Theta}
=
(\mathbf{f}, \mathbf{s}, \boldsymbol{\delta}, \boldsymbol{\pi}, \tau, \gamma)
$ denote the collection of all parameters in the model. For any scalar, vector, or block $\eta$, $\boldsymbol{\Theta}_{-\eta}$ denotes the collection of all remaining unknown parameters except $\eta$. Each iteration proceeds as follows:

\begin{enumerate}[label=\textbf{Step \arabic*.}, leftmargin=*]
    \item \textbf{Underlying signal ($\mathbf{f}$):} The full conditional distribution of $\mathbf{f}$ is multivariate normal. To facilitate efficient sampling, we work in the graph spectral domain by applying the graph Fourier transform (GFT). Specifically, let $\tilde{\mathbf f} = \mathbf U^\top \mathbf f$ denote the graph Fourier coefficients.  We first sample $\tilde{\mathbf{f}}$ from
    \[
    \tilde{\mathbf{f}} \mid \boldsymbol{\Theta}_{-\mathbf{f}},\bfy \sim \mathcal{N}\left(
    (\tau\mathbf{I}_N+\gamma\boldsymbol{\Lambda})^{-1} \tau \mathbf{U}^\top \mathbf{y}^{\mathrm{eff}}, \;
    (\tau\mathbf{I}_N+\gamma\boldsymbol{\Lambda})^{-1}
    \right),
    \]
    and then obtain $\mathbf f$ via the inverse graph Fourier transform (IGFT), $\mathbf f = \mathbf U \tilde{\mathbf f}$.

    \item \textbf{Outlier indicators and magnitudes ($s_i$ and $\delta_i$):} For each node $i=1,\dots,N$, define the residual $r_i=y_i-f_i$. We first sample the binary indicator $s_i$ conditional on the current $\delta_i$:
    \[
    s_i \mid \boldsymbol{\Theta}_{-s_i},\bfy \sim \mathrm{Bernoulli}(p_i), \qquad \log\frac{p_i}{1-p_i} = \log\frac{\pi_i}{1-\pi_i} + \tau \delta_i r_i - \frac{\tau}{2}\delta_i^2.
    \]
    Then, conditional on the updated $s_i$, we sample the slab variable $\delta_i$:
    \[
    \delta_i \mid \boldsymbol{\Theta}_{-\delta_i},\bfy \sim \mathcal{N}\left( \frac{s_i \tau r_i}{\tau_\delta + s_i \tau}, \; (\tau_\delta + s_i \tau)^{-1} \right).
    \]

    \item \textbf{Outlier probabilities ($\pi_i$):} For each $i=1,\dots,N$, we sample $\pi_i$ from its conjugate Beta distribution:
    \[
    \pi_i \mid \boldsymbol{\Theta}_{-\pi_i},\bfy \sim \mathrm{Beta}(\alpha+s_i, \beta+1-s_i).
    \]

    \item \textbf{Precision parameters ($\tau$ and $\gamma$):} We sample the observation precision $\tau$ and the graph regularization parameter $\gamma$ from their conjugate Gamma distributions. Note that $r=\mathrm{rank}(\mathbf{L})$.
    \[
    \tau \mid \boldsymbol{\Theta}_{-\tau},\bfy \sim \mathrm{Gamma}\left( \frac{N-1}{2}, \frac{1}{2} \left\| \mathbf{y}-\mathbf{f}-\mathbf{s}\odot\boldsymbol{\delta} \right\|_2^2 \right),
    \]
    \[
    \gamma \mid \boldsymbol{\Theta}_{-\gamma},\bfy \sim \mathrm{Gamma}\left( \frac{r-1}{2}, \frac{1}{2}\mathbf{f}^\top \mathbf{L}\mathbf{f} \right).
    \]
\end{enumerate}

After discarding burn-in samples, posterior summaries are obtained from the retained draws. For each node $i$, the posterior probability of being an outlier, $P(s_i=1 \mid \bf y)$, is estimated empirically as the proportion of posterior samples in which the indicator variable $s_i=1$. Node $i$ is classified as an outlier if this probability exceeds 0.5.

\section{Simulation Studies}\label{sec:simul_studies}

\subsection{Simulation Setup} \label{sec:setup}
In this section, we evaluate the performance of the proposed outlier detection method using synthetic datasets generated on various graph structures. We consider two types of random graphs: the Erd\H{o}s--R\'enyi model \citep{erdds1959random} and random geometric graphs, the latter of which is constructed based on the Euclidean distance between points uniformly distributed on a unit square. Additionally, we utilize the United States (US) sensor network \citep{kim2025quantile}, where nodes represent weather stations located near major US cities.

For a given graph Laplacian $\mathbf{L}$, a smooth underlying signal $\mathbf{f}$ is generated as follows:
\begin{equation*}
\mathbf{f} = (\mathbf{L} + 0.1 \mathbf{I})^{-1/2} \mathbf{z}, \qquad \mathbf{z} \sim \mathcal{N}(0, \mathbf{I}). \label{eq:signal_gen}
\end{equation*}
This generation process is motivated by a Gaussian Markov Random Field (GMRF) with a precision matrix $\mathbf{L}$, where the signal can be represented as $\mathbf{L}^{-1/2} \mathbf{z}$. We add $0.1 \mathbf{I}$ to the Laplacian to ensure invertibility and avoid singularity issues.

To construct the observed signal, we first add Gaussian noise $\epsilon \sim \mathcal{N}(0, \operatorname{Var}(\mathbf{f})/2)$ to the smooth signal $\mathbf{f}$, which corresponds to a signal-to-noise ratio (SNR) of 2. While the main results focus on this setting, additional experiments with $\text{SNR}\in\{1,4\}$ are provided in Appendix~\ref{sec:additional}. Subsequently, a subset of nodes is randomly selected as outliers, and outlier magnitudes drawn from a normal distribution with a mean of $ \max_i |f_i|$ and a standard deviation of $\mathrm{sd}(\mathbf{f})/2$ are added to the noisy signal. Each outlier magnitude is assigned a positive or negative sign with equal probability. This setup ensures that the outlying values are clearly distinguishable from the noisy signal.

We repeat the simulation for 100 iterations, varying the set of outliers in each instance, and report the average performance across various methods. Given that outlier detection is inherently an imbalanced classification problem, we evaluate the performance using the $F_1$ score, recall, and precision as primary metrics. In addition, we use the area under the curve (AUC) for comparison, as the choice of threshold for each method can influence outlier detection results.

We compare the proposed method with three distinct baseline approaches. To provide a comprehensive evaluation, we consider both general-purpose outlier detection methods adapted to graph signals and graph-specific methods. First, we consider Isolation Forest \citep{liu2008isolation}, which explicitly isolates anomalies for multivariate data through recursive partitioning.
Second, we consider a local median filtering (LMF) approach that combines neighborhood median smoothing with MAD-based outlier detection. Specifically, for a given graph signal $\bf y$, the local baseline $\hat{\bff}$ is obtained by taking, at each node, the median of the signal values over its neighborhood, including the node itself. Residuals are then computed as $\bf y - \hat{\bff}$, and outlier nodes are identified using the MAD-based detection method of \cite{iglewicz1993volume}. 
Third, we employ a vertex-frequency analysis approach \citep{lewenfus2019use}, reviewed in Section~\ref{sec:relatedwork}, which extracts localized features using the spectral graph wavelet transform (SGWT) and subsequently applies the local outlier factor (LOF) algorithm. 

We apply Isolation Forest only to graph settings with spatial coordinates, namely the random geometric graphs and the US sensor network. In these cases, each node is represented by a three-dimensional feature vector consisting of its two spatial coordinates and the observed signal value. For graphs without coordinate information, such as the Erd\H{o}s--R'enyi model, Isolation Forest is not applied, as it does not naturally incorporate graph topology.

\subsection{Results} \label{sec:simul_res}

\subsubsection{Erd\H{o}s--R\'enyi Model} \label{sec:erdos_renyi}
We construct Erd\H{o}s--R\'enyi random graphs consisting of $100$ nodes, where each edge is independently included with edge probability $p \in \{0.3, 0.5, 0.7\}$. As the graph is unweighted, the adjacency matrix itself serves as the weighted adjacency matrix $\mathbf{W}$.

Following the data generation procedure described in Section~\ref{sec:setup}, we generate synthetic signals across 100 independent trials for each value of $p$, with five nodes randomly selected as outliers. The results are summarized in Table~\ref{tab:outlier_performance_by_p}. 

The proposed method consistently outperforms or performs comparably to the baselines across different edge probabilities. In particular, it achieves the highest $F_1$ score, recall, and precision in most cases, while exhibiting comparable AUC values to the competing methods.

\begin{table}[t]
\centering
\caption{Performance comparison of outlier detection methods on the Erd\H{o}s--R\'enyi random graph with edge probability $p$. The values represent the mean and standard deviation (in parentheses) for each metric.}
\resizebox{\textwidth}{!}{%
\begin{tabular}{l c c c c c}
\toprule
Method & $p$ & $F_1$ & Recall & Precision & AUC \\
\midrule
Local Median Filtering & \multirow{3}{*}{0.3} & 0.672 (0.221) & 0.577 (0.259) & 0.905 (0.170) & 0.953 (0.078) \\
SGWT+LOF               &                      & 0.357 (0.157) & 0.600 (0.264) & 0.260 (0.123) & 0.842 (0.128) \\
Proposed               &                      & 0.704 (0.223) & 0.609 (0.269) & 0.936 (0.141) & 0.948 (0.078) \\
\midrule
Local Median Filtering & \multirow{3}{*}{0.5} & 0.683 (0.228) & 0.607 (0.277) & 0.895 (0.161) & 0.962 (0.059) \\
SGWT+LOF               &                      & 0.395 (0.152) & 0.645 (0.265) & 0.293 (0.123) & 0.863 (0.114) \\
Proposed               &                      & 0.662 (0.237) & 0.570 (0.284) & 0.921 (0.143) & 0.953 (0.075) \\
\midrule
Local Median Filtering & \multirow{3}{*}{0.7} & 0.716 (0.237) & 0.639 (0.294) & 0.936 (0.135) & 0.971 (0.049) \\
SGWT+LOF               &                      & 0.465 (0.196) & 0.677 (0.299) & 0.371 (0.173) & 0.869 (0.127) \\
Proposed               &                      & 0.751 (0.221) & 0.684 (0.280) & 0.923 (0.142) & 0.970 (0.056) \\
\bottomrule
\end{tabular}
}
\label{tab:outlier_performance_by_p}
\end{table}

\subsubsection{Random Geometric Graph} \label{sec:random_geometric}
We generate 100 points uniformly on the unit square and construct a random geometric graph by connecting pairs of points whose Euclidean distance is less than $0.3$. For the connected nodes $i$ and $j$, the edge weight is defined as $w_{ij}=\exp (-d_{ij}^2 / {\bar{d}}^2 )$, where $d_{ij}$ denotes the Euclidean distance between the two nodes and $\bar d$ is the average edge length over all edges in the graph. 

Following the same experimental setup, we select five nodes as outliers in each iteration. The results are summarized in Table~\ref{tab:random_geometric}. Similar to the results in the Erd\H{o}s--R\'enyi model, the proposed method demonstrates competitive performance against the baselines, particularly achieving the highest $F_1$ score, while maintaining comparable AUC values to the competing methods.

\begin{table}[t]
\centering
\caption{Performance comparison of outlier detection methods on the random geometric graph. The values represent the mean and standard deviation (in parentheses) for each metric.}
\begin{tabular}{l c c c c}
\toprule
Method & $F_1$ & Recall & Precision & AUC \\
\midrule
Isolation Forest       & 0.499 (0.159) & 0.725 (0.210) & 0.386 (0.136) & 0.916 (0.080) \\
Local Median Filtering & 0.690 (0.193) & 0.657 (0.266) & 0.831 (0.187) & 0.960 (0.055) \\
SGWT+LOF               & 0.483 (0.157) & 0.756 (0.251) & 0.382 (0.161) & 0.918 (0.091) \\
Proposed               & 0.707 (0.218) & 0.632 (0.276) & 0.908 (0.151) & 0.959 (0.059) \\
\bottomrule
\end{tabular}%
\label{tab:random_geometric}
\end{table}

\subsubsection{US Sensor Network} \label{sec:us_sensor}
The US sensor network dataset consists of 218 nodes representing weather stations. The graph is constructed using the $k$-nearest neighbors ($k$-NN) method with $k=7$, and the edge weights are assigned following the same procedure described in Section~\ref{sec:random_geometric}. For each simulation iteration, we randomly select 10 nodes as outliers.

The experimental results are summarized in Table~\ref{tbl:us_results}. Our proposed method demonstrates competitive performance compared to all considered baselines, including Isolation Forest, local median filtering, and the SGWT+LOF approach. In particular, the proposed method achieves the best performance in terms of the $F_1$ score, precision, and AUC. These results confirm that the proposed approach remains effective and robust when applied to real-world network structures, consistently surpassing the benchmark models across primary evaluation metrics.

\begin{table}[!t]
\centering
\caption{Performance comparison of outlier detection methods on the US sensor network. The values represent the mean and standard deviation (in parentheses) for each metric.}
\begin{tabular}{lcccc}
\toprule
Method & $F_1$ score & Recall & Precision & AUC \\
\midrule
Isolation Forest       & 0.473 (0.114) & 0.593 (0.150) & 0.398 (0.102) & 0.893 (0.063) \\
Local Median Filtering & 0.645 (0.147) & 0.619 (0.200) & 0.725 (0.171) & 0.949 (0.047) \\
SGWT+LOF               & 0.546 (0.126) & 0.669 (0.181) & 0.479 (0.129) & 0.878 (0.088) \\
Proposed               & 0.683 (0.162) & 0.603 (0.207) & 0.845 (0.146) & 0.955 (0.042) \\
\bottomrule
\end{tabular}
\label{tbl:us_results}
\end{table}

\begin{remark}[Analysis of SGWT+LOF Performance]
    The vertex-frequency analysis approach of \cite{lewenfus2019use} performs relatively poorly across all metrics and experimental settings considered in this study. We attribute this performance gap to the following factors. First, the discriminative power of SGWT critically depends on the choice of wavelet scale and the mother kernel. Although \cite{lewenfus2019use} suggests that different kernels may perform well at different scales, selecting an appropriate scale in practice is nontrivial, especially in the absence of prior knowledge about the spectral characteristics of outliers. This may lead to suboptimal feature representations. Second, in certain graph structures, eigenvectors may not be strongly localized, which can cause SGWT atoms to spread information across nodes. As a result, the local density contrast that LOF relies on may be attenuated, making it more difficult to distinguish outlying nodes from normal ones.
\end{remark}

\begin{remark}[Uncertainty Quantification via Posterior Samples]
     A key advantage of the proposed Bayesian framework is its ability to provide a probabilistic assessment of each node's outlier status. Using posterior samples, we compute the outlier probability for each node empirically as the proportion of samples in which the indicator variable $s_i$ equals 1. Figure~\ref{fig:us_sensor_plots} illustrates a representative outlier detection result from a specific simulation iteration on the US sensor network. As shown in the right panel of Figure~\ref{fig:us_sensor_plots}, nodes are classified as outliers if the posterior probability of being an outlier exceeds 0.5, while the continuous color scale simultaneously visualizes the degree of outlierness across nodes. This ability to quantify uncertainty represents a key strength of our method, offering more comprehensive inferential insights than deterministic baseline approaches.
\end{remark}

\begin{figure}[!t]
    \centering
    \includegraphics[width=0.98\linewidth]{figures/us_sensor_plots.png}
    \caption{Visualization of outlier detection on the US sensor network: a graph signal with 10 ground-truth outliers highlighted in magenta (left) and detection results of the proposed method, where node colors represent the posterior outlier probabilities ranging from blue (normal) to red (outlier) (right).}
    \label{fig:us_sensor_plots}
\end{figure}

\begin{remark}[Practical Limitations of AUC]
    Although the proposed method does not always achieve the highest AUC, it is important to note that AUC evaluates the ability to distinguish between outliers and normal nodes across all possible thresholds and therefore does not directly reflect performance under a specific binary decision rule. In practice, however, outlier detection requires selecting a threshold to classify nodes as outliers. For competing methods that produce only relative outlier scores, such a threshold is not always straightforward to determine. By contrast, the proposed Bayesian method provides posterior probabilities of being an outlier, for which 0.5 serves as a natural default cutoff. Using this threshold, the proposed method often achieves a higher $F_1$ score than competing methods, highlighting its practical advantage.
\end{remark}

Overall, the proposed method demonstrates competitive performance across all evaluated metrics on both randomly generated and real-world graphs, while additionally providing uncertainty quantification. 

\section{Real Data Analysis}\label{sec:real_data}
We apply the proposed model to perform outlier detection on PM2.5 data from monitoring stations in California. The data are obtained from the U.S. Environmental Protection Agency (EPA) \citep{epa2024pm25}. When a station is identified as an outlier, it may reflect either a potential sensor malfunction, which cannot be directly verified, or the influence of external factors. In this study, we focus on the latter possibility and explore the relationship between detected outliers and wildfire occurrences. Indeed, \cite{burke2023contribution} show that wildfires can substantially affect local PM2.5 levels. 

Wildfire data for California are obtained from the California Natural Resources Agency (CNRA) \citep{cnra2024wildfire}. According to these data, there were 548 fires exceeding 5,000 acres in 2024, of which 371 occurred between May and July, indicating that wildfire activity is concentrated during this period. Accordingly, we focus on PM2.5 data from this period and examine outliers on three representative dates: May 15, June 15, and July 15. These detected outliers are then compared with wildfires that were active or occurred on the corresponding dates.

\begin{figure}
    \centering
    \includegraphics[width=0.4\linewidth]{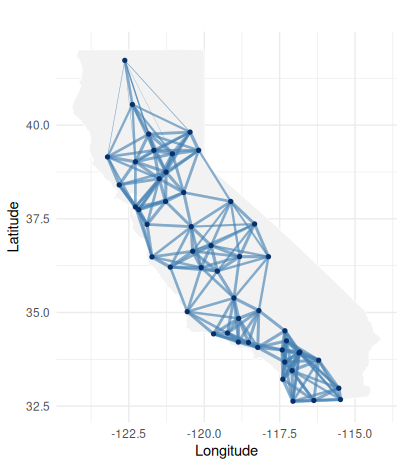}
    \caption{Visualization of the graph constructed from the 50 selected monitoring stations.}
    \label{fig:weigh_newtork_pm}
\end{figure}

For computational convenience, we select 50 out of the 159 monitoring stations. Specifically, we apply $K$-means clustering and choose one representative station from each cluster to represent the corresponding local region. To incorporate the spatial structure, we construct a graph using the $k$-NN method with $k=7$, and assign edge weights following the procedure described in Section~\ref{sec:random_geometric}. Figure~\ref{fig:weigh_newtork_pm} visualizes the resulting graph, where thicker edges indicate larger weights.

\begin{figure}[htbp]
    \centering
    \includegraphics[width=1\linewidth]{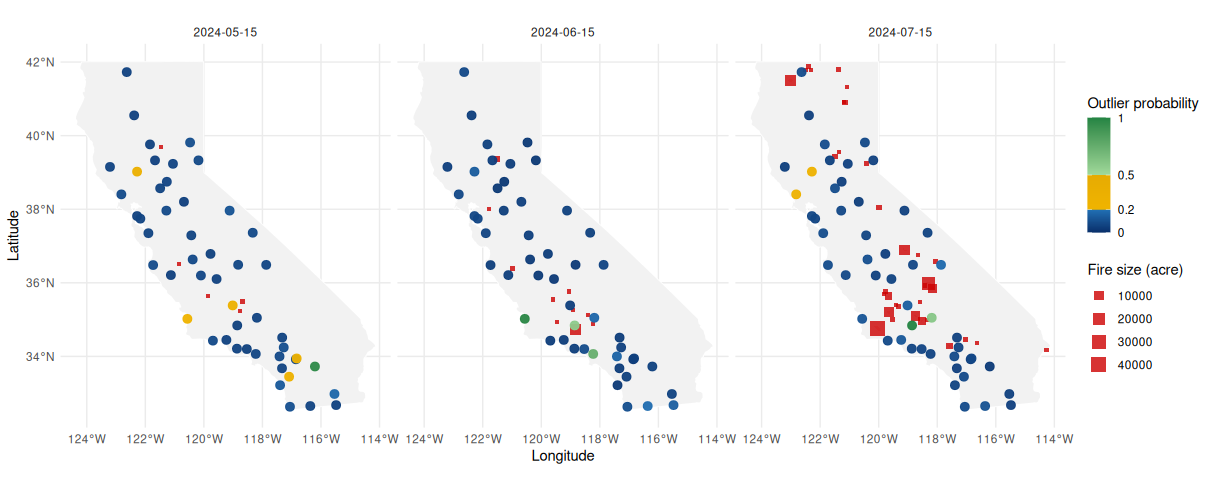}
    \caption{Daily PM2.5 outlier detection results and wildfire occurrences. Station color indicates the posterior probability of being an outlier: blue for low $[0,0.2]$, yellow for moderate $(0.2,0.5]$, and green for high $(0.5,1]$. Red squares indicate wildfire burned area, with size proportional to acres burned.}
    \label{fig:daily_outlier_pm}
\end{figure}

Figure~\ref{fig:daily_outlier_pm} illustrates the outlier detection results for May 15, June 15, and July 15, along with the corresponding wildfire data. The red square markers indicating wildfires are scaled up for visual clarity, as the actual burned areas are relatively small compared to the entire state of California. On these dates, wildfire events are observed mainly around latitude 35$^\circ$N and longitude 119$^\circ$W, and some stations with high posterior probabilities of being outliers are located near these areas. This pattern is particularly noticeable on July 15, when wildfire activity is more extensive and several nearby stations exhibit elevated outlier probabilities. Although the spatial correspondence is not exact, the results indicate a meaningful association between detected outliers and wildfire-affected regions.

\section{Conclusion}\label{sec:conclusion}
In this paper, we propose a Bayesian framework for node-level outlier detection on graph signals, which quantifies the posterior probability of each node being an outlier. We develop an efficient Gibbs sampling algorithm for posterior inference and demonstrate that the proposed method achieves competitive or superior performance compared to existing baselines across various graph structures. In particular, it often attains the highest $F_1$ score while maintaining strong performance across other evaluation metrics. We further illustrate the practical utility of the proposed approach through an application to PM2.5 data, where detected outliers are interpreted in relation to wildfire occurrences.

We conclude with several directions for future work. While the current study focuses on graph signals observed at a single time point, an important extension is to settings where graph signals are observed over time. In such cases, we may also consider incorporating dynamic graph structures. Moreover, the proposed framework could be extended to incorporate additional sources of information, such as covariates or more complex dependency structures.

\section*{Acknowledgements}
This research was supported by the National Research Foundation of Korea (NRF) funded by the Korea government (MSIT) (RS-2026-25471513), by the Basic Science Research Program through the NRF grant funded by the Ministry of Education (RS-2021-NR060140), and by Global - Learning \& Academic research institution for Master's$\cdot$PhD students, and Postdocs(G-LAMP) Program of the NRF grant funded by the Ministry of Education(No. RS-2025-25442252).

\clearpage

\begin{appendices}
\section{Proof of Theorem~\ref{thm:post_proper}} \label{sec:proofs}
\begin{proof}
We assume that the observed signal satisfies $\bar{\bf y} \neq 0$ and $y_i \neq 0$ for all $i=1,\ldots,N$. These conditions exclude only a Lebesgue-null set of degenerate observations.

Let
\[
\mathcal{S}=\{0,1\}^N,
\qquad
\mathbf D_{\mathbf s}=\mathrm{diag}(s_1,\ldots,s_N).
\]
Since the graph is connected, the graph Laplacian has the spectral decomposition
\[
\mathbf L = \mathbf U_r \boldsymbol{\Lambda} \mathbf U_r^\top,
\]
where $\mathbf U_r\in\mathbb R^{N\times r}$ has orthonormal columns spanning the subspace orthogonal to the null space of $\mathbf L$, and
\[
\boldsymbol{\Lambda}=\mathrm{diag}(\lambda_1,\ldots,\lambda_r),
\qquad \lambda_j>0,\quad r=N-1.
\]
Let
\[
\mathbf u_0 = N^{-1/2}\mathbf 1_N,
\]
so that $(\mathbf u_0,\mathbf U_r)$ is an orthogonal basis of $\mathbb R^N$.

Since the IGMRF prior is understood on the subspace orthogonal to $\ker(\mathbf L)$, we write
\[
\mathbf f = \mathbf U_r \mathbf z,
\qquad \mathbf z\in\mathbb R^r.
\]
Then, the prior for $\mathbf z$ is
\[
\pi(\mathbf z\mid \gamma)
\propto
\gamma^{r/2}\exp\!\left(-\frac{\gamma}{2}\mathbf z^\top \boldsymbol{\Lambda}\mathbf z\right).
\]

\paragraph{Step 1: Reduction to finitely many terms.}
Because $\alpha,\beta>0$, the Beta prior for each $\pi_i$ is proper. Hence, for any fixed $\mathbf s\in\mathcal S$,
\[
\int_0^1 \pi_i^{\alpha+s_i-1}(1-\pi_i)^{\beta+(1-s_i)-1}\,d\pi_i
=
B(\alpha+s_i,\beta+1-s_i)<\infty,
\]
where $B(a,b)= \dfrac{\Gamma(a)\Gamma(b)}{\Gamma(a+b)}$ for $a,b>0$, and $\Gamma(\cdot)$ denotes the Gamma function.

Therefore, after integrating out $\boldsymbol{\pi}$, the posterior normalizing constant can be written as a finite sum
\[
Z = \sum_{\mathbf s\in\mathcal S} c_{\mathbf s}\, I_{\mathbf s},
\]
where
\[
c_{\mathbf s}
=
\prod_{i=1}^N B(\alpha+s_i,\beta+1-s_i)
\in (0,\infty),
\]
and
\[
I_{\mathbf s}
=
\int_{\mathbb R^N}
\int_0^\infty
\int_0^\infty
\int_{\mathbb R^r}
p(\mathbf y\mid \mathbf z,\mathbf s,\boldsymbol{\delta},\tau)\,
\pi(\mathbf z\mid \gamma)\,
\pi(\boldsymbol{\delta})\,
\pi(\tau)\pi(\gamma)
\,d\mathbf z\,d\gamma\,d\tau\,d\boldsymbol{\delta}.
\]
Since the number of states $\mathbf s\in\mathcal S$ is finite, it suffices to show that $I_{\mathbf s}<\infty$ for every $\mathbf s$.

\paragraph{Step 2: Integrating out the latent graph signal.}
Fix $\mathbf s\in\mathcal S$ and $\boldsymbol{\delta}\in\mathbb R^N$, and define
\[
\mathbf w_{\mathbf s}(\boldsymbol{\delta})
:=
\mathbf y-\mathbf D_{\mathbf s}\boldsymbol{\delta}.
\]
Since $(\mathbf u_0,\mathbf U_r)$ is an orthogonal basis of $\mathbb R^N$, write
\[
a_{\mathbf s}(\boldsymbol{\delta})
:=
\mathbf u_0^\top \mathbf w_{\mathbf s}(\boldsymbol{\delta}),
\qquad
\mathbf b_{\mathbf s}(\boldsymbol{\delta})
:=
\mathbf U_r^\top \mathbf w_{\mathbf s}(\boldsymbol{\delta})
=
\bigl(b_{\mathbf s1}(\boldsymbol{\delta}),\ldots,b_{\mathbf sr}(\boldsymbol{\delta})\bigr)^\top.
\]
Then,
\[
\mathbf w_{\mathbf s}(\boldsymbol{\delta})
=
a_{\mathbf s}(\boldsymbol{\delta})\,\mathbf u_0
+
\mathbf U_r \mathbf b_{\mathbf s}(\boldsymbol{\delta}),
\]
and, recalling that $\mathbf f=\mathbf U_r\mathbf z$,
\[
\mathbf y-\mathbf f-\mathbf D_{\mathbf s}\boldsymbol{\delta}
=
a_{\mathbf s}(\boldsymbol{\delta})\,\mathbf u_0
+
\mathbf U_r\bigl(\mathbf b_{\mathbf s}(\boldsymbol{\delta})-\mathbf z\bigr).
\]
By orthogonality,
\[
\|\mathbf y-\mathbf f-\mathbf D_{\mathbf s}\boldsymbol{\delta}\|^2
=
a_{\mathbf s}(\boldsymbol{\delta})^2
+
\|\mathbf b_{\mathbf s}(\boldsymbol{\delta})-\mathbf z\|^2.
\]

Hence, for fixed $(\mathbf s,\boldsymbol{\delta},\tau,\gamma)$,
\begin{align*}
p(\mathbf y\mid \mathbf z,\mathbf s,\boldsymbol{\delta},\tau)\,
\pi(\mathbf z\mid\gamma)
\propto\;&
\tau^{N/2}\gamma^{r/2}
\exp\!\left(-\frac{\tau}{2}a_{\mathbf s}(\boldsymbol{\delta})^2\right) \\
&\times
\exp\!\left(
-\frac12\sum_{j=1}^r
\Bigl[
\tau\bigl(z_j-b_{\mathbf sj}(\boldsymbol{\delta})\bigr)^2
+
\gamma\lambda_j z_j^2
\Bigr]
\right).
\end{align*}
For each $j=1,\ldots,r$, completing the square gives
\[
\tau\bigl(z_j-b_{\mathbf sj}(\boldsymbol{\delta})\bigr)^2
+
\gamma\lambda_j z_j^2
=
(\tau+\gamma\lambda_j)
\left(
z_j-\frac{\tau\,b_{\mathbf sj}(\boldsymbol{\delta})}{\tau+\gamma\lambda_j}
\right)^2
+
\frac{\tau\gamma\lambda_j}{\tau+\gamma\lambda_j}
\,b_{\mathbf sj}(\boldsymbol{\delta})^2.
\]
Including the factors from $\pi(\tau)\pi(\gamma)$ and integrating out
$\mathbf z=(z_1,\ldots,z_r)$ coordinatewise yields
\begin{align}
I_{\mathbf s}
&\propto
\int_{\mathbb R^N}\phi_{\tau_\delta}(\boldsymbol{\delta})
\int_0^\infty\int_0^\infty
K_{\mathbf s}(\tau,\gamma,\boldsymbol{\delta})
\,d\gamma\,d\tau\,d\boldsymbol{\delta},
\label{eq:Is_after_z}
\end{align}
where
\begin{align}
K_{\mathbf s}(\tau,\gamma,\boldsymbol{\delta})
:=
\tau^{(r-2)/2}\gamma^{(r-3)/2}
\prod_{j=1}^r (\tau+\gamma\lambda_j)^{-1/2}
\exp\!\left(
-\frac{\tau}{2}a_{\mathbf s}(\boldsymbol{\delta})^2
-\frac12\sum_{j=1}^r
\frac{\tau\gamma\lambda_j}{\tau+\gamma\lambda_j}
\,b_{\mathbf sj}(\boldsymbol{\delta})^2
\right),
\label{eq:K_s_def}
\end{align}
and $\phi_{\tau_\delta}(\boldsymbol{\delta})$ denotes the joint Gaussian density of
$\boldsymbol{\delta}\sim\mathcal N(\mathbf 0,\tau_\delta^{-1}\mathbf I_N)$.

Equation \eqref{eq:Is_after_z}, together with the kernel definition
\eqref{eq:K_s_def}, will be the starting point for the remainder of the proof.
\paragraph{Step 3: Integrability in the region $0<\tau\le 1$.}
We first show that the contribution from the region $0<\tau\le 1$ is finite, uniformly in $\mathbf s$ and $\boldsymbol{\delta}$.

Since the exponential factor in \eqref{eq:K_s_def} is bounded above by $1$, we have
\[
K_{\mathbf s}(\tau,\gamma,\boldsymbol{\delta})
\le
\tau^{(r-2)/2}\gamma^{(r-3)/2}
\prod_{j=1}^r (\tau+\gamma\lambda_j)^{-1/2}.
\]
Hence
\[
\int_0^\infty K_{\mathbf s}(\tau,\gamma,\boldsymbol{\delta})\,d\gamma
\le
\tau^{(r-2)/2}
\int_0^\infty
\gamma^{(r-3)/2}
\prod_{j=1}^r (\tau+\gamma\lambda_j)^{-1/2}\,d\gamma.
\]
Define
\[
J(\tau)
:=
\int_0^\infty
\gamma^{(r-3)/2}
\prod_{j=1}^r (\tau+\gamma\lambda_j)^{-1/2}\,d\gamma.
\]
Using the change of variables $\gamma=\tau u$, we obtain
\begin{align*}
J(\tau)
&=
\int_0^\infty
(\tau u)^{(r-3)/2}
\prod_{j=1}^r \bigl(\tau+\tau u\lambda_j\bigr)^{-1/2}
\,\tau\,du \\
&=
\tau^{(r-3)/2}\tau^{-r/2}\tau
\int_0^\infty
u^{(r-3)/2}
\prod_{j=1}^r (1+u\lambda_j)^{-1/2}\,du \\
&=
\tau^{-1/2}
\int_0^\infty
u^{(r-3)/2}
\prod_{j=1}^r (1+u\lambda_j)^{-1/2}\,du.
\end{align*}
Let
\[
C_\Lambda
:=
\int_0^\infty
u^{(r-3)/2}
\prod_{j=1}^r (1+u\lambda_j)^{-1/2}\,du.
\]
We now verify that $C_\Lambda<\infty$.

For $0<u\le 1$, since each factor satisfies
\[
0<(1+u\lambda_j)^{-1/2}\le 1,
\]
we have
\[
u^{(r-3)/2}\prod_{j=1}^r (1+u\lambda_j)^{-1/2}
\le
u^{(r-3)/2}.
\]
Because $r>1$, the exponent $(r-3)/2$ is strictly greater than $-1$, and therefore
\[
\int_0^1
u^{(r-3)/2}\,du
<\infty.
\]

For $u\ge 1$, we use
\[
(1+u\lambda_j)^{-1/2}\le (\lambda_j u)^{-1/2}
= \lambda_j^{-1/2}u^{-1/2},
\]
so that
\[
u^{(r-3)/2}\prod_{j=1}^r (1+u\lambda_j)^{-1/2}
\le
\Bigl(\prod_{j=1}^r \lambda_j^{-1/2}\Bigr)
u^{(r-3)/2}u^{-r/2}
=
C_1\,u^{-3/2},
\]
where $C_1 = \prod_{j=1}^r \lambda_j^{-1/2}$ is a positive constant independent of $u$.

Since
\[
\int_1^\infty u^{-3/2}\,du<\infty,
\]
it follows that $C_\Lambda<\infty$.

Therefore,
\[
J(\tau)=C_\Lambda\tau^{-1/2},
\]
and hence
\[
\int_0^\infty K_{\mathbf s}(\tau,\gamma,\boldsymbol{\delta})\,d\gamma
\le
C_\Lambda \tau^{(r-2)/2}\tau^{-1/2}
=
C_\Lambda \tau^{(r-3)/2}.
\]
It follows that
\[
\int_0^1\int_0^\infty
K_{\mathbf s}(\tau,\gamma,\boldsymbol{\delta})
\,d\gamma\,d\tau
\le
C_\Lambda
\int_0^1 \tau^{(r-3)/2}\,d\tau
<\infty,
\]
because $r>1$. Since this bound does not depend on $\boldsymbol{\delta}$ and the Gaussian prior on $\boldsymbol{\delta}$ is proper, the contribution from the region $0<\tau\le 1$ is finite.

\paragraph{Step 4: The case $\mathbf s=\mathbf 0$.}
When $\mathbf s=\mathbf 0$, the likelihood does not depend on $\boldsymbol{\delta}$, and
\[
a_{\mathbf 0}(\boldsymbol{\delta})
=
\mathbf u_0^\top \mathbf y
=
\sqrt N\,\bar y, \quad \bar y := \dfrac{1}{N}\sum_{i=1}^N y_i,
\]
which is nonzero by assumption. Hence the exponential factor in \eqref{eq:K_s_def} contains
\[
\exp\!\left(-\frac{N\bar y^2}{2}\tau\right).
\]
Using the bound from Step 3,
\[
\int_0^\infty K_{\mathbf 0}(\tau,\gamma,\boldsymbol{\delta})\,d\gamma
\le
C_\Lambda \tau^{(r-3)/2}\exp\!\left(-\frac{N\bar y^2}{2}\tau\right).
\]
Therefore,
\[
\int_1^\infty\int_0^\infty
K_{\mathbf 0}(\tau,\gamma,\boldsymbol{\delta})
\,d\gamma\,d\tau
\le
C_\Lambda\int_1^\infty
\tau^{(r-3)/2}\exp\!\left(-\frac{N\bar y^2}{2}\tau\right)
\,d\tau
<\infty.
\]
Since the prior on $\boldsymbol{\delta}$ is proper, this implies $I_{\mathbf 0}<\infty$.

\paragraph{Step 5: The mixed cases $0<|\mathbf s|<N$.}
Fix $\mathbf s$ such that $0<|\mathbf s|<N$. Then, there exists at least one inactive coordinate $i_0$ such that $s_{i_0}=0$. By assumption, $y_{i_0}\neq 0$. Since the $i_0$-th coordinate is inactive,
\[
\bigl(\mathbf y-\mathbf D_{\mathbf s}\boldsymbol{\delta}\bigr)_{i_0}=y_{i_0}
\qquad\text{for all }\boldsymbol{\delta}.
\]
Write
\[
\mathbf w_{\mathbf s}(\boldsymbol{\delta})
=
\mathbf y-\mathbf D_{\mathbf s}\boldsymbol{\delta}.
\]
Because
\[
\mathbf w_{\mathbf s}(\boldsymbol{\delta})
=
a_{\mathbf s}(\boldsymbol{\delta})\mathbf u_0
+
\mathbf U_r\mathbf b_{\mathbf s}(\boldsymbol{\delta}),
\]
we have
\[
\|\mathbf w_{\mathbf s}(\boldsymbol{\delta})\|^2
=
a_{\mathbf s}(\boldsymbol{\delta})^2
+
\|\mathbf b_{\mathbf s}(\boldsymbol{\delta})\|^2.
\]
Moreover,
\[
\|\mathbf w_{\mathbf s}(\boldsymbol{\delta})\|
\ge |y_{i_0}|.
\]
Choose $\eta_{\mathbf s}\in(0,|y_{i_0}|)$ and define
\[
A_{\mathbf s}
=
\{\boldsymbol{\delta}: |a_{\mathbf s}(\boldsymbol{\delta})|>\eta_{\mathbf s}\},
\qquad
B_{\mathbf s}
=
\{\boldsymbol{\delta}: |a_{\mathbf s}(\boldsymbol{\delta})|\le \eta_{\mathbf s}\}.
\]
Then, for $\boldsymbol{\delta}\in B_{\mathbf s}$,
\[
\|\mathbf b_{\mathbf s}(\boldsymbol{\delta})\|^2
=
\|\mathbf w_{\mathbf s}(\boldsymbol{\delta})\|^2
-
a_{\mathbf s}(\boldsymbol{\delta})^2
\ge
y_{i_0}^2-\eta_{\mathbf s}^2
=: \rho_{\mathbf s}^2>0.
\]

\medskip
\noindent
\emph{Contribution over $A_{\mathbf s}$.}
For $\boldsymbol{\delta} \in A_{\mathbf s}$, we have
\[
\exp\!\left(-\frac{\tau}{2}a_{\mathbf s}(\boldsymbol{\delta})^2\right)
\le
\exp\!\left(-\frac{\eta_{\mathbf s}^2}{2}\tau\right).
\]
Hence, using the bound from Step 3,
\[
\int_1^\infty \int_0^\infty
K_{\mathbf s}(\tau,\gamma,\boldsymbol{\delta})
\,d\gamma\,d\tau
\le
C_\Lambda
\int_1^\infty
\tau^{(r-3)/2}
\exp\!\left(-\frac{\eta_{\mathbf s}^2}{2}\tau\right)
\,d\tau
<\infty.
\]
Therefore the contribution from $A_{\mathbf s}$ is finite.

\medskip
\noindent
\emph{Contribution over $B_{\mathbf s}$.}
For $\boldsymbol{\delta}\in B_{\mathbf s}$ and $\tau\ge 1$, split the $\gamma$-integral into two parts:
\[
\int_0^\infty = \int_0^\tau + \int_\tau^\infty.
\]

If $0<\gamma\le \tau$, then
\[
\tau+\gamma\lambda_j \ge \tau,
\qquad
\frac{\tau\gamma\lambda_j}{\tau+\gamma\lambda_j}
\ge
\frac{\gamma\lambda_j}{1+\lambda_j}
\ge c_1\gamma
\]
for some constant $c_1>0$ depending only on $\{\lambda_j\}_{j=1}^r$. Since
\[
\|\mathbf b_{\mathbf s}(\boldsymbol{\delta})\|^2\ge \rho_{\mathbf s}^2,
\]
we obtain
\begin{align}
\int_0^\tau
\gamma^{(r-3)/2}
\prod_{j=1}^r(\tau+\gamma\lambda_j)^{-1/2}
\exp\!\left(
-\frac12\sum_{j=1}^r \frac{\tau\gamma\lambda_j}{\tau+\gamma\lambda_j}
\,b_{\mathbf sj}(\boldsymbol{\delta})^2
\right)
\,d\gamma
&\le
\tau^{-r/2}
\int_0^\infty
\gamma^{(r-3)/2}
e^{-c_1\rho_{\mathbf s}^2\gamma/2}\,d\gamma \notag\\
&\le
C_{\mathbf s,1}\tau^{-r/2},
\label{eq:step5_integral_bound1}
\end{align}
for some finite constant $C_{\mathbf s,1}>0$ depending only on $\mathbf s$.

If $\gamma>\tau$, then
\[
\tau+\gamma\lambda_j \ge \gamma\lambda_j,
\qquad
\frac{\tau\gamma\lambda_j}{\tau+\gamma\lambda_j}
\ge
\frac{\tau\lambda_j}{1+\lambda_j}
\ge c_2\tau
\]
for some constant $c_2>0$. Therefore,
\begin{align}
\int_\tau^\infty
\gamma^{(r-3)/2}
\prod_{j=1}^r(\tau+\gamma\lambda_j)^{-1/2}
\exp\!\left(
-\frac12\sum_{j=1}^r \frac{\tau\gamma\lambda_j}{\tau+\gamma\lambda_j}
\,b_{\mathbf sj}(\boldsymbol{\delta})^2
\right)
\,d\gamma
&\le
C_1
e^{-c_2\rho_{\mathbf s}^2\tau/2}
\int_\tau^\infty \gamma^{-3/2}\,d\gamma \notag\\
&=
2C_1\tau^{-1/2}e^{-c_2\rho_{\mathbf s}^2\tau/2} \notag\\
&\le
C_{\mathbf s,2}\tau^{-r/2},
\label{eq:step5_integral_bound2}
\end{align}
where
\[
C_{\mathbf s,2}
:=
2C_1\sup_{\tau\ge 1}\tau^{(r-1)/2}e^{-c_2\rho_{\mathbf s}^2\tau/2}
<\infty.
\]

Combining \eqref{eq:step5_integral_bound1} and \eqref{eq:step5_integral_bound2}, we obtain
\[
\int_0^\infty
\gamma^{(r-3)/2}
\prod_{j=1}^r(\tau+\gamma\lambda_j)^{-1/2}
\exp\!\left(
-\frac12\sum_{j=1}^r \frac{\tau\gamma\lambda_j}{\tau+\gamma\lambda_j}
\,b_{\mathbf sj}(\boldsymbol{\delta})^2
\right)
\,d\gamma
\le
C_{\mathbf s,3}\tau^{-r/2},
\]
where $C_{\mathbf s,3} := \max(C_{\mathbf s,1},C_{\mathbf s,2})$.

Substituting this into \eqref{eq:K_s_def}, we obtain
\[
\int_0^\infty K_{\mathbf s}(\tau,\gamma,\boldsymbol{\delta})\,d\gamma
\le
C_{\mathbf s,3}\tau^{-1}
\exp\!\left(-\frac{\tau}{2}a_{\mathbf s}(\boldsymbol{\delta})^2\right),
\qquad \tau\ge 1,\ \boldsymbol{\delta}\in B_{\mathbf s}.
\]

Now
\[
a_{\mathbf s}(\boldsymbol{\delta})
=
\mathbf u_0^\top \mathbf y
-
(\mathbf u_0\odot \mathbf s)^\top \boldsymbol{\delta}.
\]
Since $\mathbf s\neq \mathbf 0$, the vector $\mathbf u_0\odot \mathbf s$ is nonzero. Under the Gaussian prior
\[
\boldsymbol{\delta}\sim \mathcal N(\mathbf 0,\tau_\delta^{-1}\mathbf I_N),
\]
the random variable
\[
M_{\mathbf s}:=(\mathbf u_0\odot \mathbf s)^\top \boldsymbol{\delta}
\]
is Gaussian with variance
\[
\sigma_{\mathbf s}^2
=
\frac{\|\mathbf u_0\odot \mathbf s\|^2}{\tau_\delta}
>0.
\]
Therefore,
\begin{align*}
\int_{\mathbb R^N}
\phi_{\tau_\delta}(\boldsymbol{\delta})
\exp\!\left(-\frac{\tau}{2}a_{\mathbf s}(\boldsymbol{\delta})^2\right)
\,d\boldsymbol{\delta}
&=
\mathbb E\!\left[
\exp\!\left(
-\frac{\tau}{2}(\mathbf u_0^\top \mathbf y - M_{\mathbf s})^2
\right)
\right]
\notag\\
&=
\frac{1}{\sqrt{1+\sigma_{\mathbf s}^2\tau}}
\exp\!\left(
-\frac{\tau(\mathbf u_0^\top \mathbf y)^2}{2(1+\sigma_{\mathbf s}^2\tau)}
\right)
\le
C_{\mathbf s}\tau^{-1/2},
\qquad \tau\ge 1.
\label{eq:delta_convolution_bound_revised}
\end{align*}
Hence,
\[
\int_{B_{\mathbf s}}\phi_{\tau_\delta}(\boldsymbol{\delta})
\int_1^\infty\int_0^\infty
K_{\mathbf s}(\tau,\gamma,\boldsymbol{\delta})
\,d\gamma\,d\tau\,d\boldsymbol{\delta}
\le
C_{\mathbf s}\int_1^\infty \tau^{-3/2}\,d\tau
<\infty.
\]
Therefore $I_{\mathbf s}<\infty$ for every $0<|\mathbf s|<N$.

\paragraph{Step 6: The case $\mathbf s=\mathbf 1_N$.}
When $\mathbf s=\mathbf 1_N$, all nodes are active, and it is convenient to integrate out
$\boldsymbol{\delta}$ first:
\[
\int
\mathcal N\!\left(
\mathbf y\mid \mathbf f+\boldsymbol{\delta},\tau^{-1}\mathbf I_N
\right)
\mathcal N\!\left(
\boldsymbol{\delta}\mid \mathbf 0,\tau_\delta^{-1}\mathbf I_N
\right)
\,d\boldsymbol{\delta}
=
\mathcal N\!\left(
\mathbf y\mid \mathbf f,\ (\tau^{-1}+\tau_\delta^{-1})\mathbf I_N
\right).
\]
Define
\[
\kappa(\tau):=\frac{\tau\tau_\delta}{\tau+\tau_\delta}.
\]
Then, $0<\kappa(\tau)\le \tau_\delta$, $\kappa(\tau)\sim \tau$ as $\tau\downarrow 0$, and
\[
\inf_{\tau\ge 1}\kappa(\tau)=\kappa(1)=\frac{\tau_\delta}{1+\tau_\delta}>0.
\]
After integrating out $\boldsymbol{\delta}$, the same calculation as in Step~2 yields a kernel of the form
\[
\widetilde K(\tau,\gamma)
=
\tau^{-3/2}\,\kappa(\tau)^{N/2}\,
\gamma^{(r-3)/2}
\prod_{j=1}^r \bigl(\kappa(\tau)+\gamma\lambda_j\bigr)^{-1/2}
\times \exp(\cdots),
\]
where the exponential factor satisfies $\exp(\cdots)\le 1$.

\medskip
\noindent
\emph{Contribution over $0<\tau\le 1$.}
For $0<\tau\le 1$, we have
\[
\frac{\tau_\delta}{\tau_\delta+1}\,\tau \le \kappa(\tau)\le \tau.
\]
Bounding $\exp(\cdots)\le 1$ and using the function $J(\cdot)$ from Step~3, we obtain
\[
\int_0^\infty \widetilde K(\tau,\gamma)\,d\gamma
\le
\tau^{-3/2}\,\kappa(\tau)^{N/2}\,J(\kappa(\tau))
=
C_{\Lambda}\,\tau^{-3/2}\,\kappa(\tau)^{(N-1)/2}.
\]
Since $\kappa(\tau)\le \tau$ and $r=N-1$, it follows that
\[
\int_0^\infty \widetilde K(\tau,\gamma)\,d\gamma
\le
C_{\Lambda}\,\tau^{-3/2}\,\tau^{(N-1)/2}
=
C_{\Lambda}\,\tau^{(r-3)/2}.
\]
Therefore,
\[
\int_0^1\int_0^\infty \widetilde K(\tau,\gamma)\,d\gamma\,d\tau
<\infty,
\]
because $r>1$.

\medskip
\noindent
\emph{Contribution over $\tau\ge 1$.}
For $\tau\ge 1$, we have $\kappa(1)\le \kappa(\tau)\le \tau_\delta$, so $\kappa(\tau)$ is bounded above and bounded away from $0$ on $[1,\infty)$.
Hence, by Step~3,
\[
J(\kappa(\tau))=C_{\Lambda}\kappa(\tau)^{-1/2}\le C<\infty
\qquad \text{uniformly for }\tau\ge 1,
\]
and also $\kappa(\tau)^{N/2}\le \tau_\delta^{N/2}$. Consequently,
\[
\int_0^\infty \widetilde K(\tau,\gamma)\,d\gamma
\le
C\,\tau^{-3/2},
\qquad \tau\ge 1,
\]
which is integrable on $[1,\infty)$. Combining the two regions yields $I_{\mathbf 1_N}<\infty$.

Overall, we have shown that $I_{\mathbf s}<\infty$ for every $\mathbf s\in\{0,1\}^N$. Since the number of such configurations is finite and the Beta integrals are finite, the posterior normalizing constant is finite. Therefore the posterior distribution is proper.

\end{proof}

\section{Additional Simulation Results} \label{sec:additional}
In this section, we report additional simulation results for signal-to-noise ratio (SNR) values of 1 and 4. Overall, the proposed method remains competitive across the considered settings, and performance generally improves as the SNR increases.
Tables~\ref{tab:outlier_performance_by_p_snr4}, \ref{tab:random_geometric_snr4}, and \ref{tbl:us_results_snr4} present the results for $\text{SNR}=4$. Tables~\ref{tab:outlier_performance_by_p_snr1}, \ref{tab:random_geometric_snr1}, and \ref{tbl:us_results_snr1} present the results for $\text{SNR}=1$.

\begin{table}[!htb]
\centering
\caption{Performance comparison of outlier detection methods on the Erd\H{o}s--R\'enyi random graph with edge probability $p$ for $\mathrm{SNR}=4$. The values represent the mean and standard deviation (in parentheses) for each metric.}
\resizebox{\textwidth}{!}{%
\begin{tabular}{l c c c c c}
\toprule
Method & $p$ & $F_1$ & Recall & Precision & AUC \\
\midrule
Local Median Filtering & \multirow{3}{*}{0.3} & 0.731 (0.219) & 0.651 (0.277) & 0.932 (0.144) & 0.959 (0.082) \\
SGWT+LOF               &                      & 0.394 (0.145) & 0.691 (0.250) & 0.280 (0.111) & 0.858 (0.124) \\
Proposed               &                      & 0.741 (0.210) & 0.658 (0.266) & 0.947 (0.108) & 0.957 (0.073) \\
\midrule
Local Median Filtering & \multirow{3}{*}{0.5} & 0.719 (0.227) & 0.647 (0.280) & 0.914 (0.147) & 0.972 (0.050) \\
SGWT+LOF               &                      & 0.412 (0.151) & 0.671 (0.256) & 0.305 (0.119) & 0.881 (0.106) \\
Proposed               &                      & 0.731 (0.232) & 0.647 (0.285) & 0.952 (0.114) & 0.966 (0.056) \\
\midrule
Local Median Filtering & \multirow{3}{*}{0.7} & 0.777 (0.215) & 0.719 (0.280) & 0.938 (0.122) & 0.979 (0.040) \\
SGWT+LOF               &                      & 0.469 (0.199) & 0.688 (0.307) & 0.378 (0.183) & 0.886 (0.123) \\
Proposed               &                      & 0.784 (0.219) & 0.724 (0.275) & 0.931 (0.129) & 0.978 (0.043) \\
\bottomrule
\end{tabular}
}
\label{tab:outlier_performance_by_p_snr4}
\end{table}

\begin{table}[!htb]
\centering
\caption{Performance comparison of outlier detection methods on the random geometric graph for $\mathrm{SNR}=4$. The values represent the mean and standard deviation (in parentheses) for each metric.}
\begin{tabular}{l c c c c}
\toprule
Method & $F_1$ & Recall & Precision & AUC \\
\midrule
Isolation Forest       & 0.521 (0.150) & 0.759 (0.198) & 0.401 (0.129) & 0.928 (0.070) \\
Local Median Filtering & 0.710 (0.188) & 0.693 (0.268) & 0.829 (0.177) & 0.966 (0.050) \\
SGWT+LOF               & 0.496 (0.151) & 0.790 (0.234) & 0.380 (0.144) & 0.930 (0.083) \\
Proposed               & 0.745 (0.198) & 0.679 (0.258) & 0.911 (0.140) & 0.968 (0.047) \\
\bottomrule
\end{tabular}
\label{tab:random_geometric_snr4}
\end{table}

\begin{table}[!htb]
\centering
\caption{Performance comparison of outlier detection methods on the US sensor network for $\mathrm{SNR}=4$. The values represent the mean and standard deviation (in parentheses) for each metric.}
\begin{tabular}{lcccc}
\toprule
Method & $F_1$ score & Recall & Precision & AUC \\
\midrule
Isolation Forest       & 0.517 (0.109) & 0.650 (0.147) & 0.433 (0.097) & 0.918 (0.054) \\
Local Median Filtering & 0.718 (0.129) & 0.751 (0.187) & 0.722 (0.148) & 0.970 (0.033) \\
SGWT+LOF               & 0.608 (0.115) & 0.770 (0.165) & 0.521 (0.132) & 0.920 (0.064) \\
Proposed               & 0.776 (0.138) & 0.744 (0.192) & 0.849 (0.128) & 0.972 (0.035) \\
\bottomrule
\end{tabular}
\label{tbl:us_results_snr4}
\end{table}

\begin{table}[!htb]
\centering
\caption{Performance comparison of outlier detection methods on the Erd\H{o}s--R\'enyi random graph with edge probability $p$ for $\mathrm{SNR}=1$. The values represent the mean and standard deviation (in parentheses) for each metric.}
\resizebox{\textwidth}{!}{%
\begin{tabular}{l c c c c c}
\toprule
Method & $p$ & $F_1$ & Recall & Precision & AUC \\
\midrule
Local Median Filtering & \multirow{3}{*}{0.3} & 0.605 (0.225) & 0.495 (0.257) & 0.909 (0.179) & 0.944 (0.070) \\
SGWT+LOF               &                      & 0.330 (0.148) & 0.560 (0.280) & 0.244 (0.114) & 0.804 (0.131) \\
Proposed               &                      & 0.608 (0.224) & 0.495 (0.251) & 0.913 (0.172) & 0.934 (0.082) \\
\midrule
Local Median Filtering & \multirow{3}{*}{0.5} & 0.580 (0.235) & 0.494 (0.288) & 0.879 (0.183) & 0.939 (0.082) \\
SGWT+LOF               &                      & 0.368 (0.152) & 0.587 (0.265) & 0.277 (0.124) & 0.823 (0.127) \\
Proposed               &                      & 0.579 (0.243) & 0.477 (0.279) & 0.893 (0.175) & 0.917 (0.110) \\
\midrule
Local Median Filtering & \multirow{3}{*}{0.7} & 0.668 (0.239) & 0.581 (0.294) & 0.926 (0.151) & 0.954 (0.073) \\
SGWT+LOF               &                      & 0.443 (0.187) & 0.633 (0.289) & 0.364 (0.172) & 0.846 (0.142) \\
Proposed               &                      & 0.674 (0.249) & 0.591 (0.294) & 0.905 (0.165) & 0.954 (0.075) \\
\bottomrule
\end{tabular}
}
\label{tab:outlier_performance_by_p_snr1}
\end{table}

\begin{table}[!htb]
\centering
\caption{Performance comparison of outlier detection methods on the random geometric graph for $\mathrm{SNR}=1$. The values represent the mean and standard deviation (in parentheses) for each metric.}
\begin{tabular}{l c c c c}
\toprule
Method & $F_1$ & Recall & Precision & AUC \\
\midrule
Isolation Forest       & 0.467 (0.169) & 0.681 (0.230) & 0.360 (0.140) & 0.893 (0.102) \\
Local Median Filtering & 0.645 (0.212) & 0.604 (0.291) & 0.833 (0.189) & 0.947 (0.070) \\
SGWT+LOF               & 0.461 (0.158) & 0.714 (0.268) & 0.362 (0.147) & 0.899 (0.107) \\
Proposed               & 0.656 (0.228) & 0.572 (0.281) & 0.895 (0.168) & 0.946 (0.071) \\
\bottomrule
\end{tabular}
\label{tab:random_geometric_snr1}
\end{table}

\begin{table}[!htb]
\centering
\caption{Performance comparison of outlier detection methods on the US sensor network for $\mathrm{SNR}=1$. The values represent the mean and standard deviation (in parentheses) for each metric.}
\begin{tabular}{lcccc}
\toprule
Method & $F_1$ score & Recall & Precision & AUC \\
\midrule
Isolation Forest       & 0.402 (0.117) & 0.498 (0.148) & 0.341 (0.106) & 0.843 (0.076) \\
Local Median Filtering & 0.527 (0.178) & 0.462 (0.214) & 0.691 (0.197) & 0.902 (0.069) \\
SGWT+LOF               & 0.455 (0.151) & 0.514 (0.203) & 0.440 (0.164) & 0.816 (0.105) \\
Proposed               & 0.541 (0.196) & 0.436 (0.213) & 0.819 (0.176) & 0.911 (0.062) \\
\bottomrule
\end{tabular}
\label{tbl:us_results_snr1}
\end{table}
\end{appendices}


\clearpage
	
\bibliographystyle{dcu}
\bibliography{signal}

@article{justel2001bayesian,
  title={Bayesian unmasking in linear models},
  author={Justel, Ana and Pe{\~n}a, Daniel},
  journal={Computational statistics \& data analysis},
  volume={36},
  number={1},
  pages={69--84},
  year={2001},
  publisher={Elsevier}
}

@article{silva2019bayesian,
  title={Bayesian outlier detection in non-Gaussian autoregressive time series},
  author={Silva, Maria Eduarda and Pereira, Isabel and McCabe, Brendan},
  journal={Journal of Time Series Analysis},
  volume={40},
  number={5},
  pages={631--648},
  year={2019},
  publisher={Wiley Online Library}
}

@article{box1968bayesian,
  title={A Bayesian approach to some outlier problems},
  author={Box, George EP and Tiao, George C},
  journal={Biometrika},
  volume={55},
  number={1},
  pages={119--129},
  year={1968},
  publisher={Oxford University Press}
}

@article{george1993variable,
  title={Variable selection via Gibbs sampling},
  author={George, Edward I and McCulloch, Robert E},
  journal={Journal of the American Statistical Association},
  volume={88},
  number={423},
  pages={881--889},
  year={1993},
  publisher={Taylor \& Francis}
}

@inproceedings{lewenfus2019use,
  title={On the use of vertex-frequency analysis for anomaly detection in graph signals},
  author={Lewenfus, Gabriela and Alves Martins, Wallace and Chatzinotas, Symeon and Ottersten, Bj{\"o}rn},
  booktitle={XXXVII Simp{\'o}sio Brasileiro de Telecomunica{\c{c}}{\~o}es e Processamento de Sinais (SBrT 2019)},
  year={2019}
}

@article{sandryhaila2014discrete,
  title={Discrete signal processing on graphs: Frequency analysis},
  author={Sandryhaila, Aliaksei and Moura, Jose MF},
  journal={IEEE Transactions on signal processing},
  volume={62},
  number={12},
  pages={3042--3054},
  year={2014},
  publisher={IEEE}
}

@article{sandryhaila2013discrete,
  title={Discrete signal processing on graphs},
  author={Sandryhaila, Aliaksei and Moura, Jos{\'e} MF},
  journal={IEEE transactions on signal processing},
  volume={61},
  number={7},
  pages={1644--1656},
  year={2013},
  publisher={IEEE}
}

@article{shuman2013emerging,
  title={The emerging field of signal processing on graphs: Extending high-dimensional data analysis to networks and other irregular domains},
  author={Shuman, David I and Narang, Sunil K and Frossard, Pascal and Ortega, Antonio and Vandergheynst, Pierre},
  journal={IEEE signal processing magazine},
  volume={30},
  number={3},
  pages={83--98},
  year={2013},
  publisher={IEEE}
}

@inproceedings{gopalakrishnan2019identification,
  title={Identification of outliers in graph signals},
  author={Gopalakrishnan, Karthik and Li, Max Z and Balakrishnan, Hamsa},
  booktitle={2019 IEEE 58th Conference on Decision and Control (CDC)},
  pages={4769--4776},
  year={2019},
  organization={IEEE}
}

@inproceedings{egilmez2014spectral,
  title={Spectral anomaly detection using graph-based filtering for wireless sensor networks},
  author={Egilmez, Hilmi E and Ortega, Antonio},
  booktitle={2014 IEEE International Conference on Acoustics, Speech and Signal Processing (ICASSP)},
  pages={1085--1089},
  year={2014},
  organization={IEEE}
}

@article{qin2026adaptive,
  title={Adaptive Graph Learning for Anomaly Detection in Wireless Sensor Networks},
  author={Qin, Junhui and He, Ning and Wang, Kuiyang and Zhao, Zhipeng and Jiang, Hongyan},
  journal={IEEE Access},
  year={2026},
  publisher={IEEE}
}

@inproceedings{shuman2012windowed,
  title={A windowed graph Fourier transform},
  author={Shuman, David I and Ricaud, Benjamin and Vandergheynst, Pierre},
  booktitle={2012 IEEE Statistical Signal Processing Workshop (SSP)},
  pages={133--136},
  year={2012},
  organization={Ieee}
}

@article{hammond2011wavelets,
  title={Wavelets on graphs via spectral graph theory},
  author={Hammond, David K and Vandergheynst, Pierre and Gribonval, R{\'e}mi},
  journal={Applied and computational harmonic analysis},
  volume={30},
  number={2},
  pages={129--150},
  year={2011},
  publisher={Elsevier}
}

@article{francisquini2022community,
  title={Community-based anomaly detection using spectral graph filtering},
  author={Francisquini, Rodrigo and Lorena, Ana Carolina and Nascimento, Mari{\'a} CV},
  journal={Applied Soft Computing},
  volume={118},
  pages={108489},
  year={2022},
  publisher={Elsevier}
}

@article{gelman2006prior,
  title={Prior distributions for variance parameters in hierarchical models},
  author={Gelman, A},
  journal={Bayesian Analysis},
  volume={1},
  pages={515--534},
  year={2006}
}

@book{rue2005gaussian,
  title={Gaussian Markov random fields: theory and applications},
  author={Rue, Havard and Held, Leonhard},
  year={2005},
  publisher={Chapman and Hall/CRC}
}

@article{burke2023contribution,
  title={The contribution of wildfire to PM2. 5 trends in the USA},
  author={Burke, Marshall and Childs, Marissa L and de la Cuesta, Brandon and Qiu, Minghao and Li, Jessica and Gould, Carlos F and Heft-Neal, Sam and Wara, Michael},
  journal={Nature},
  volume={622},
  number={7984},
  pages={761--766},
  year={2023},
  publisher={Nature Publishing Group UK London}
}

@article{erdds1959random,
  title={On random graphs I},
  author={Erd\H{o}s, P and R\'enyi, A},
  journal={Publ. math. debrecen},
  volume={6},
  number={290-297},
  pages={18},
  year={1959}
}

@article{kim2025quantile,
  title={Quantile-based fitting for graph signals},
  author={Kim, Kyusoon and Oh, Hee-Seok},
  journal={Statistics and Computing},
  volume={35},
  number={6},
  pages={180},
  year={2025},
  publisher={Springer}
}

@inproceedings{liu2008isolation,
  title={Isolation forest},
  author={Liu, Fei Tony and Ting, Kai Ming and Zhou, Zhi-Hua},
  booktitle={2008 eighth ieee international conference on data mining},
  pages={413--422},
  year={2008},
  organization={IEEE}
}

@inproceedings{breunig2000lof,
  title={LOF: identifying density-based local outliers},
  author={Breunig, Markus M and Kriegel, Hans-Peter and Ng, Raymond T and Sander, J{\"o}rg},
  booktitle={Proceedings of the 2000 ACM SIGMOD international conference on Management of data},
  pages={93--104},
  year={2000}
}

@book{iglewicz1993volume,
  title={Volume 16: how to detect and handle outliers},
  author={Iglewicz, Boris and Hoaglin, David C},
  year={1993},
  publisher={Quality Press}
}

@article{leus2023graph,
  title={Graph signal processing: History, development, impact, and outlook},
  author={Leus, Geert and Marques, Antonio G and Moura, Jos{\'e} MF and Ortega, Antonio and Shuman, David I},
  journal={IEEE Signal Processing Magazine},
  volume={40},
  number={4},
  pages={49--60},
  year={2023},
  publisher={IEEE}
}

@misc{epa2024pm25,
  author       = {{U.S. Environmental Protection Agency (EPA)}},
  title        = {Outdoor Air Quality Data: Download Daily Data},
  year         = {2024},
  url          = {https://www.epa.gov/outdoor-air-quality-data/download-daily-data},
  note         = {Accessed: 2026-04-13}
}

@misc{cnra2024wildfire,
  author       = {{California Natural Resources Agency (CNRA)}},
  title        = {California Fire Perimeters (All)},
  year         = {2024},
  url          = {https://data.cnra.ca.gov/dataset/california-fire-perimeters-all},
  note         = {Accessed: 2026-04-13}
}

\end{document}